\documentclass[a4paper,11pt]{article}
\usepackage{pos}
\usepackage[utf8]{inputenc}
\usepackage{graphicx}
\usepackage{xcolor}
\usepackage{bm}
\usepackage{hyperref}
\hypersetup{colorlinks=true,linkbordercolor=blue,linkcolor=blue, citecolor=blue,pdfborderstyle={/S/U/W 1}}
\usepackage{xspace}
\usepackage{booktabs}
\usepackage{wrapfig}
\usepackage{caption}

\newcommand{\la}[1]{\label{#1}}
\def\be{\begin{equation}}
\def\ee{\end{equation}}
\def\bea{\begin{eqnarray}}
\def\eea{\end{eqnarray}}
\newcommand{\eq}[1]{Eq.~(\ref{#1})}
\newcommand{\dd}{{\rm d}}

\newcommand{\pt}{\ensuremath{p_{_\perp}}}

\newcommand{\RpA}{\ensuremath{R_{\textnormal{pA}}}\xspace}
\newcommand{\RpPb}{\ensuremath{R_{\textnormal{pPb}}}\xspace}

\newcommand{\R}{{\textnormal{R}}\xspace}
\newcommand{\x}{\ensuremath{\mathnormal{x}}\xspace}
\newcommand{\Phat}{{\cal P}}
\newcommand{\p}{\textnormal{p}}
\newcommand{\pp}{\textnormal{pp}}
\newcommand{\pA}{\textnormal{pA}}
\newcommand{\bi}{\begin{itemize}}
\newcommand{\ei}{\end{itemize}}

\title{Probing fully coherent radiation and parton densities using (virtual) photons at the LHC}
\author*[a]{François Arleo}
\author[a]{Djessy Bourgeais}
\author[a]{Maxime Guilbaud}
\author[a]{Greg Jackson}
\author[a]{Víctor Valencia Torres}
\affiliation[a]{SUBATECH UMR 6457 (IMT Atlantique, Université de Nantes, IN2P3/CNRS)\\4 rue Alfred Kastler, 44307 Nantes, France}
\emailAdd{francois.arleo@subatech.in2p3.fr}
\emailAdd{djessy.bourgeais@subatech.in2p3.fr}
\emailAdd{maxime.guilbaud@subatech.in2p3.fr}
\emailAdd{jackson@subatech.in2p3.fr}
\emailAdd{victor.valencia@subatech.in2p3.fr}
\abstract{
Prompt photon production in pA collisions has 
long been suggested as a sensitive probe of 
the nuclear parton distribution functions (nPDFs). 
In this study, we present recent results on another 
cold nuclear matter effect, namely 
fully coherent radiation induced by parton multiple scattering, 
which may influence the nuclear dependence of prompt photon production. 
Medium-induced radiation effects, 
implemented in leading-order direct and fragmentation photon processes, 
are computed for pPb collisions at the LHC. 
At backward rapidity, 
photons are sensitive to fully coherent energy loss (FCEL), 
while at forward rapidity, 
fully coherent energy gain (FCEG) plays a crucial role 
due to the dominance of 
the $qg \to q\gamma$ scattering channel. 
In contrast, for virtual photon production, the impact 
of fully coherent radiation is marginal, 
making Drell-Yan (DY) one of the best ways to probe nuclear PDFs. 
The power of the DY process is demonstrated by reweighting 
nPDF sets at next-to-leading order  
using realistic pseudo-data for LHC Run 3.
}
\FullConference{The European Physical Society Conference on High Energy Physics (EPS-HEP2025)\\
7-11 July 2025\\
Marseille, France\\}
\begin{document}
\maketitle

\section{Introduction}\label{sec:introduction}

Electroweak probes are excellent tools 
in high energy proton-nucleus (pA) collisions, 
with the potential to distinguish  
genuine modifications of nuclear parton distribution functions (nPDFs)
from other 
cold nuclear matter 
effects. 
Among the latter,  
{\em fully coherent energy loss} 
(FCEL)\,---\,arising 
when the emission of a
soft gluon 
has a formation time 
longer than the nuclear path length\,---\,is a robust 
prediction of QCD~\cite{Arleo:2010rb}. 
Implications of FCEL 
have been studied for a variety of hard processes, 
including quarkonia, 
light- and heavy-meson production, 
and even atmospheric neutrinos~\cite{Arleo:2012hn,Arleo:2020eia,Arleo:2020hat,Arleo:2021bpv,Arleo:2021krm}.

Because FCEL scales with the energy of the incoming parton, 
it cannot be neglected and may interfere with clean 
extractions of nPDFs, 
especially if those include hadron production data  (e.g. charmonium or $D$-meson~\cite{Eskola:2021nhw,AbdulKhalek:2022fyi,Duwentaster:2022kpv}) in their global fit analyses~\cite{Arleo:2025oos}. 
This issue is particularly relevant 
at LHC energies, where energy loss competes with nPDF effects. 
In contrast, prompt photons are known to undergo  
minimal final-state interactions making them a cleaner 
experimental probe of nPDFs~\cite{Arleo:2007js,Arleo:2011gc,Helenius:2014qla}.
However, their leading-order (LO) production mechanism involves 
a recoiling parton, leaving some room 
for FCEL (and, as we shall see, 
{\em fully coherent energy gain}, FCEG) to affect photon yields 
in pA collisions. 

In these proceedings, we investigate the role of medium-induced 
coherent energy loss/gain on prompt photon production at the LHC, 
via the nuclear modification factor, 
\be\label{eq:RpA}
  R_{\pA} (y, p_\perp)
  \; \equiv \; 
  \frac{1}{A}
  \frac{\dd \sigma_{\pA}}{\dd y \, \dd p_\perp}
  \bigg/
  \frac{\dd \sigma_{\pp}}{\dd y \, \dd p_\perp}
  \; ,
\ee
highlighting the balance of colour charge at 
forward/backward rapidities (Sec.~\ref{sec:fcel}). 
We also examine the Drell-Yan (DY) process 
(Sec.~\ref{sec:npdfs}), 
which is expected to 
be far less sensitive to coherent radiation owing to 
the colourless final state (at leading order) 
and to the large invariant mass~\cite{Arleo:2015qiv}. 
Using realistic pseudo-data samples, 
and Bayesian reweighting techniques, 
we demonstrate the strong constraining power that 
future LHCb DY measurements could bring to the nPDF programme.

\section{Fully coherent radiation effects on prompt photons}\label{sec:fcel}

\begin{figure}[htbp]
    \centering
    \begin{minipage}[t]{0.47\linewidth}
        \centering
        \includegraphics[width=1.125\linewidth]{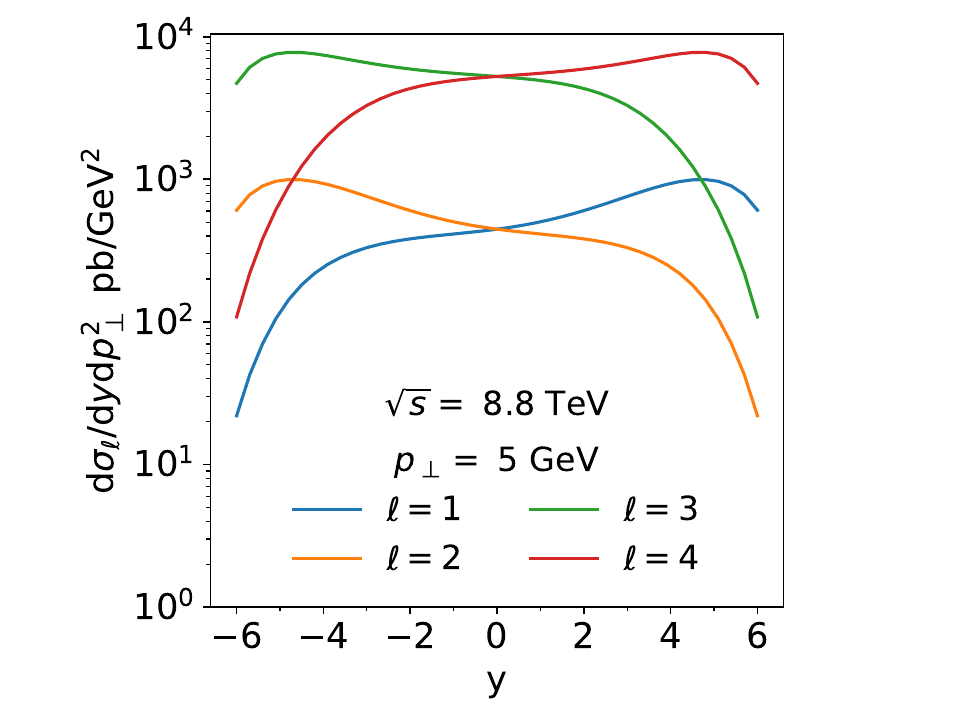}   
        \caption{Prompt photon differential cross section as a function of $y$ at fixed $\pt=5$~GeV, and for each individual partonic sub-processes.}
        \label{fig:xs}
    \end{minipage}
    \hfill 
    \begin{minipage}[t]{0.47\linewidth}
        \centering
        \includegraphics[width=0.9\linewidth]{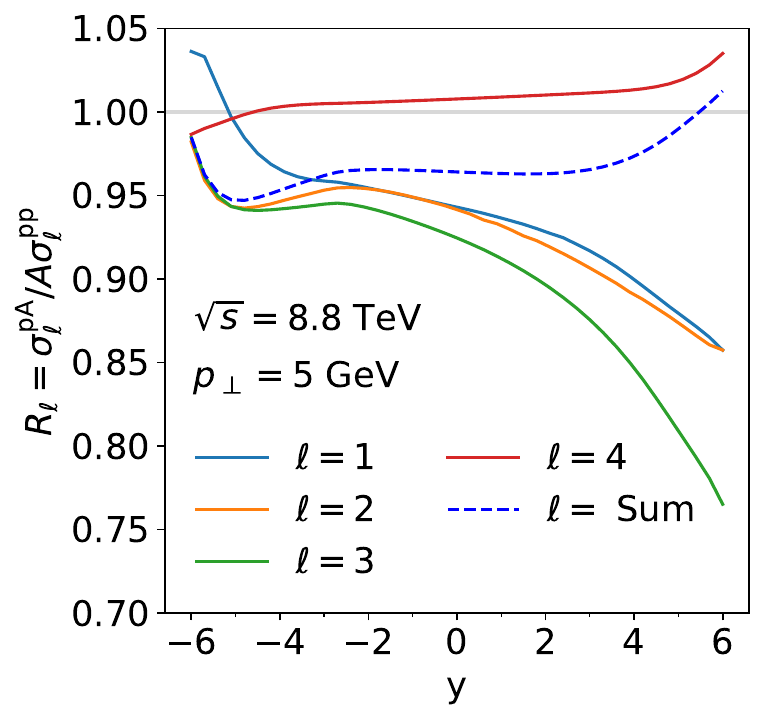}
        \caption{$\RpPb(y, \pt=5\,\textnormal{GeV})$ for the partonic sub-processes $\ell \in \{1,2,3,4\}$ and when all processes are added (dashed line).}
        \label{fig:RpA_channels}
    \end{minipage}
\end{figure}

In this section we discuss the effects of 
fully coherent radiation on prompt photon production. 
The starting point is the calculation 
of the direct photon production cross section 
in pp collisions~\cite{Aurenche:2006vj},
\be\label{eq:ppxs}
  \frac{\dd {\sigma}^{ }_{\pp}}{\dd y\,\dd\pt^2} 
  \; = \; 
  \sum_{\ell}\, 
  \frac{
    \dd {\sigma}^{\ell}_{\pp}
  }{
    \dd y\,\dd\pt^2
  } 
  \; = \;
  \sum_{\ell}\,
  \int_{\xi_{\textnormal{min}}}^{\xi_{\textnormal{max}}} 
  \dd\xi\,
  \frac{
    \dd {\sigma}^{\ell}_{\pp}
  }{
    \dd y\,\dd\pt^2\,\dd\xi
  }
  \; , 
\ee
where the sum runs over the various partonic sub-processes 
$\ell=\left\{ij\to \gamma k\right\}$ at LO 
in the strong coupling constant, 
namely Compton scattering and $q\bar{q}$ annihilation, 
listed in Table~\ref{tab:processes}. 
We introduce the variable $\xi$, 
defined as $\xi=1/\big(1+\exp(y_k-y)\big)$ 
where $y_k$ is the rapidity of the recoiling parton, 
which is bounded by 
$\xi_{\textnormal{min}} \equiv \pt e^y/\sqrt{s}$ 
and 
$\xi_{\textnormal{max}}=1-\pt e^{-y}/\sqrt{s}\,$. 
The differential cross section 
of each sub-process appearing in \eqref{eq:ppxs} is given by
\be\label{eq:ppxs_ell}
  \frac{\dd {\sigma}^{\ell}_{\pp}}{\dd y\,\dd\pt^2\,\dd\xi} 
  \; = \;
  \frac{1}{s}\,\frac{1}{\xi(1-\xi)} \, 
  f_i^\textnormal{\,p}(x_1, \mu^2) \, 
  f_j^\textnormal{\,p}(x_2, \mu^2) \,
  \,\frac{\dd {\hat{\sigma}}_{ }^{\ell}}{\dd \xi}\,,
\ee
\begin{table}
    \centering
    \vspace{-10pt}
    \small
    \begin{tabular}{cccc}       
        \toprule
        $\ell$ & Process & \(\R\) & $C_{\ell}$ \\
        \midrule
        1 & \(q\bar{q} \to g\gamma\) & \textbf{8} & \(N_c\) \\
        2 & \(\bar{q}q \to g\gamma\) & \textbf{8} & \(N_c\) \\
        3 & \(gq \to q\gamma\) & \textbf{3} & \(N_c\) \\
        4 & \(qg \to q\gamma\) & \textbf{3} & \(-1/N_c\) \\
        \bottomrule
    \end{tabular}
    \caption{LO processes for direct $\gamma$ production.}
    \label{tab:processes}
    \vspace{-5pt}
\end{table}
where the PDFs $f^\textnormal{p}$ 
are evaluated respectively at 
$x_1 = {p_\perp e^y}/\,(\xi\sqrt{s})$ 
and 
$x_2={p_\perp e^{-y}}/\,((1-\xi)\sqrt{s})$, 
the factorization scale $\mu$ is varied in the range 
$[p_\perp/2;2p_\perp]$, 
and 
$\dd\hat{\sigma}^\ell/\dd\xi$ 
is the partonic cross section of process $\ell$. 
The cross section in pp collisions at $\sqrt{s}=8.8$~TeV as a function of $y$ at fixed $\pt=5$~GeV, together with the individual sub-processes, is shown in~Figure~\ref{fig:xs}.

The effects of fully coherent radiation on 
the differential production cross section in 
pA collisions can be modeled as\footnote{More generally, the pp cross section in the r.h.s. of~\eqref{eq:pAxs} can be replaced by a incoherent sum of pp and pn collisions (responsible for isospin effects at backward rapidity~\cite{Arleo:2015qiv}), or using genuine nuclear PDF. Here we focus solely on fully coherent radiation, hence the choice of using the pp cross section in~\eqref{eq:pAxs}.}~\cite{Arleo:2020hat}
\be\label{eq:pAxs}
  \frac{1}{A}\,\frac{
    \dd {\sigma}^{ }_{\pA}(y, \pt) 
  }{
    \dd y\,\dd\pt^2
  }  
  \; = \;
  \sum_{\ell}\,
  \int_{\xi_{\textnormal{min}}}^{\xi_{\textnormal{max}}} \dd\xi\,
  \int_0^{\x_{_{\textnormal{max}}}} \! \! \dd x \, \, 
  \frac{{\Phat_{\ell}}(x, \xi)}{\left(1+x\right)^{s_\ell}} \, 
  \frac{
    \dd {\sigma}^{\ell}_{\pp} (y+s_{\ell}\ln(1+x), \pt, \xi)
  }{
    \dd y\,\dd\pt^2\,\dd\xi
  } \,,
\ee
where $s_\ell \equiv \textnormal{sign}(C_{\ell})$ 
and $C_{\ell}$ is a color factor depending on 
the Casimir of the initial-state partons and that of 
the irreducible representation \R of the final state,\footnote{%
  Since photons are color neutral, 
  there is only one color representation for each sub-process, 
  i.e. $C_\R=C_k$.
} 
$C_{\ell}=C_i+C_{\R}-C_j$. 
The sign of $C_\ell$ dictates whether 
the final state experiences fully coherent energy loss (FCEL), 
$C_\ell>0$, or fully coherent energy gain (FCEG), 
$C_\ell<0$. As can be seen in Table~\ref{tab:processes}, 
the $\ell=3$ sub-process, which dominates the cross section 
at negative rapidity, is sensitive to FCEL 
(as well as the subdominant $\ell=1,2$ channels), 
while the $\ell=4$ sub-process, dominant at positive rapidity, 
is sensitive to FCEG.

The quenching weight $\Phat_{\ell}$ appearing in 
\eq{eq:pAxs} can be approximated by~\cite{Arleo:2012rs}
\begin{equation*}
  \Phat^{ }_{{\ell}}(x,\xi) 
  \; \simeq \; 
  \left|\frac{\dd I_\ell(x,\xi)}{\dd x}\right| \, 
  \exp \left\{ \,
    - \int_{x}^{\infty} \!\! \dd x^\prime\, 
    \left|\frac{\dd I_\ell(x^\prime,\xi)}{\dd x}\right| 
  \, \right\}\,,
\end{equation*}
where the medium-induced gluon radiation spectrum 
is taken as~\cite{Jackson:2023adv,in-prep} 
\be\label{eq:spectrum}
  x\, \frac{
    \dd I_{{\ell}}(x,\xi)
  }{
  \dd x}
  \; \simeq \;
  C_{\ell}\,
  \frac{\alpha_s}{\pi}\, \left[ 
  \ln\left( 1 + \frac{(1-\xi)^2\,Q_{s}^2}{x^2\,\pt^2} \right) 
  - 
  \ln\left( 1 + \frac{(1-\xi)^2\,Q_{s,\p}^2}{x^2\,\pt^2} \right)
  \right]\,,
\ee
with $Q_s$ (resp. $Q_{s,\p}$) being the saturation scale in the nucleus 
(resp. in the proton).\footnote{%
  The expressions used for $Q_s$ and 
  $\x_{_{\textnormal{max}}}$ can be found e.g. in Ref.~\cite{Arleo:2020hat}.
} 

Using \eq{eq:pAxs} and \eq{eq:ppxs} allows 
for computing the nuclear modification factor 
$\RpA$, \eq{eq:RpA}, for each individual sub-process and for the sum. 
As shown in Figure~\ref{fig:RpA_channels}, 
the shape of $\RpPb$ (blue dashed line) results from 
the interplay of FCEL effects, 
which tend to suppress the direct photon yield at large negative $y$ 
(due to sub-processes $\ell\leq3$, which individual \RpPb's are shown as blue, orange, green lines), 
while  FCEG (due to $\ell=4$, red line) leads to a slight enhancement, $\RpPb\gtrsim1$, at large positive $y$.

\begin{figure}[htbp]
    \centering
    \includegraphics[width=0.407\linewidth]{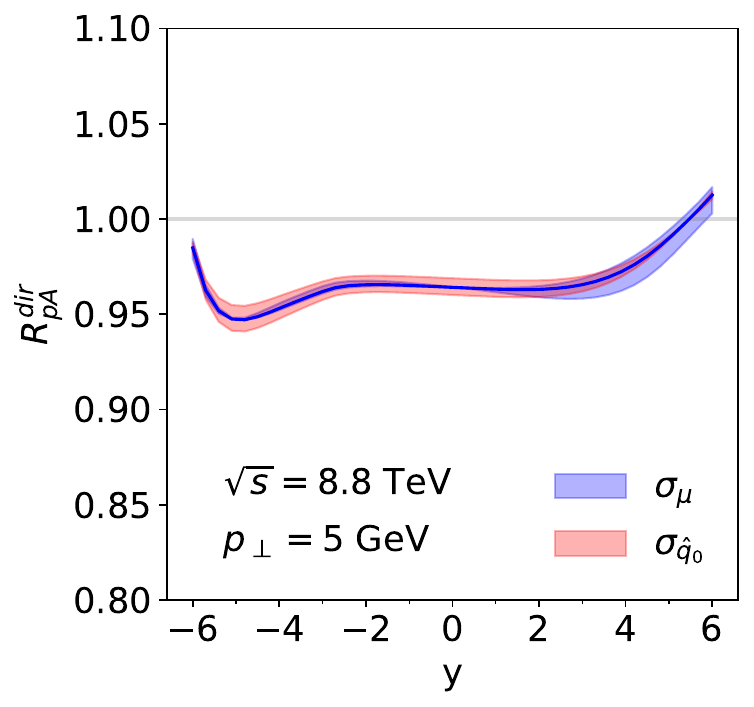}
    \hspace{3mm}
    \includegraphics[width=.517\linewidth]{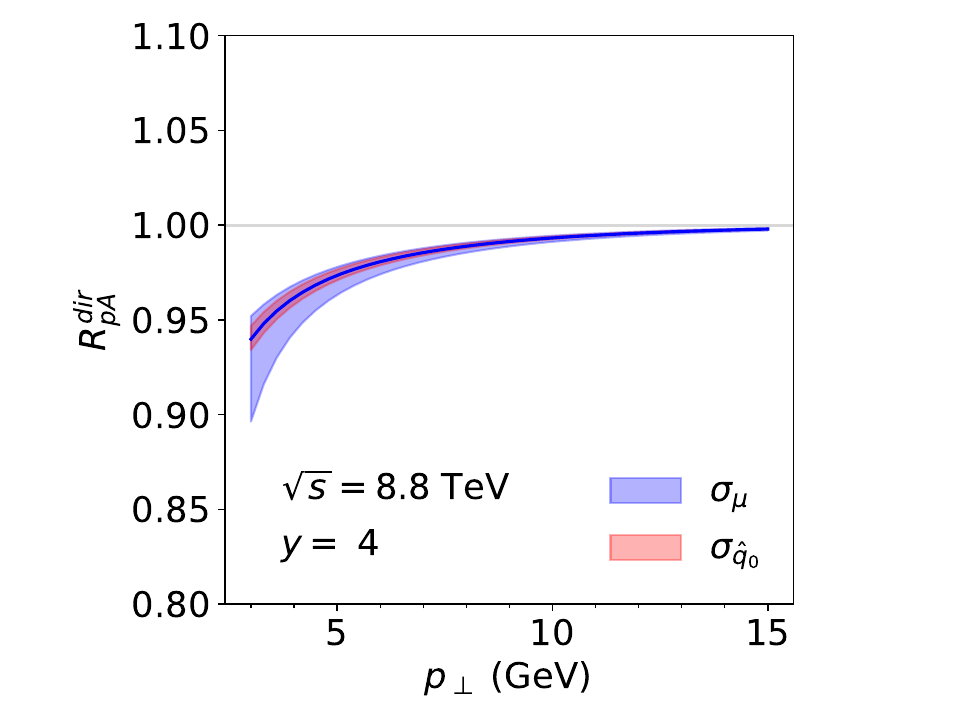}
    \caption{$\RpPb$ of direct photon as a function of $y$ at $\pt=5$~GeV (left) and as a function of $\pt$ at $y=4$ (right). The uncertainties reflect the variation of the transport coefficient and that of the factorization scale.
    }
    \label{fig:RpA}
\end{figure}

Figure~\ref{fig:RpA} (left) displays the rapidity dependence of 
$\RpPb(y, \pt)$ at fixed $\pt=5$~GeV together with 
the theoretical uncertainty. 
It arises mostly from the value 
of the cold nuclear matter transport coefficient $\hat{q}$,
which sets the magnitude of the saturation scale $Q_s\,$. 
The factorization scale dependence in the PDF, 
see~\eqref{eq:ppxs_ell}, affects the relative yields of 
partonic sub-processes\footnote{%
  The distinction between various partonic sub-processes 
  is scale-dependent and ambiguous beyond the leading order.
} 
sensitive to FCEL and FCEG and hence contributes to the uncertainty 
of $\RpPb$, mostly at large positive rapidity. 
The $\pt$ dependence of $\RpPb(y, \pt)$ at fixed $y=4$ is shown 
in Figure~\ref{fig:RpA} (right). 
As expected, $\RpPb$ reaches unity at large $\pt$ as 
the induced gluon spectrum \eqref{eq:spectrum} drops rapidly 
when $\pt \gg Q_s$. 
At low $\pt=3$~GeV, however, 
the suppression of direct photons lie in the range 
$0.90 \lesssim \RpPb \lesssim 0.95$.

\begin{figure}[htbp]
    \centering
    \includegraphics[width=0.37\linewidth]{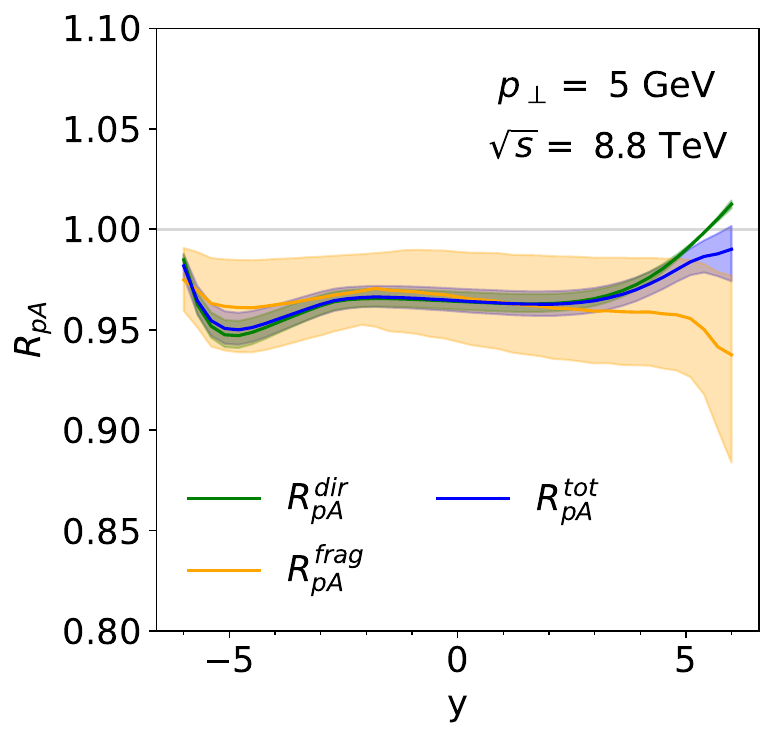}
    \hspace{0.09\linewidth}
    \includegraphics[width=.37\linewidth]{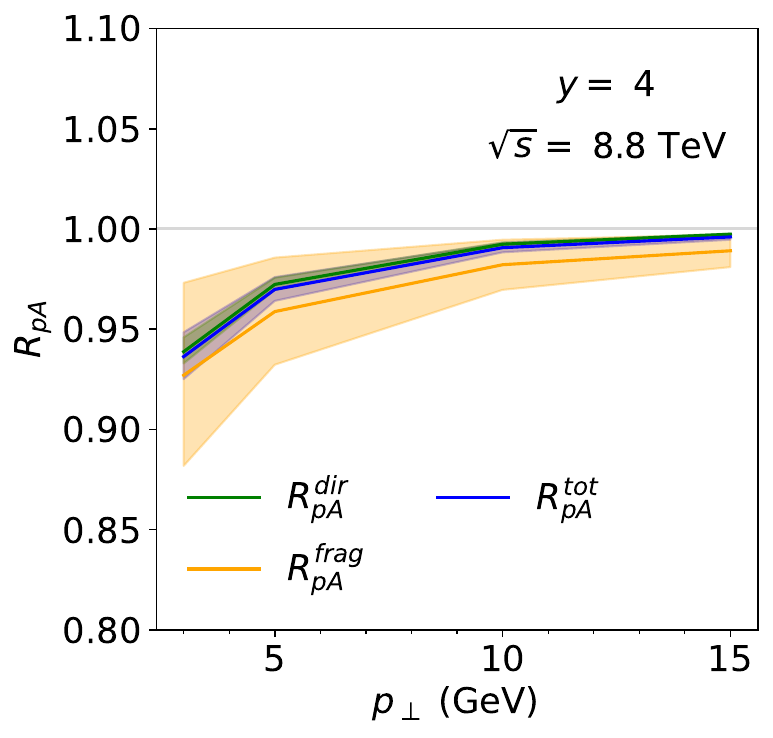}
    \caption{%
    $\RpPb$ of direct (green band), fragmentation (yellow band) and total prompt photons (blue band) as a function of $y$ at $\pt=5$~GeV (left) and as a function of $\pt$ at $y=4$ (right).
    }
    \label{fig:RpA_dir+frag}
\end{figure}

In addition to direct processes, 
prompt photons can also be produced from the fragmentation of 
hard partons in QCD processes, $i\,j\to k\,l(\to\gamma)$. 
Because of the parametric dependence of the fragmentation function, 
$D_l^\gamma={\cal O}\left(\alpha/\alpha_s\right)$, 
fragmentation photons contribute in principle to the same order 
as direct photons in perturbation theory. 
Their contribution, however, 
is strongly suppressed due to isolation requirements, 
used in the experiments to suppress the background from 
neutral meson decays (in what follows the amount of 
hadronic energy deposited within the cone of radius $R=0.4$ 
centered around the photon is required to be less than 
$10\%$ of the photon $\pt$).
The FCEL/G effects on fragmentation photons have 
been computed systematically along the lines of 
Ref.~\cite{Arleo:2020hat} on light hadron production. 
The relative yields of all partonic sub-processes and the shape of 
their rapidity distributions were evaluated using the 
\texttt{JETPHOX} Monte-Carlo generator~\cite{Aurenche:2006vj}.
The resulting $\RpPb$, together with those from direct photons 
and the sum of both components is shown in 
Figure~\ref{fig:RpA_dir+frag}. 
Although the theoretical uncertainty on $\RpPb$ is slightly enlarged, 
the inclusion of fragmentation photons does 
not dramatically affect its magnitude.

\section{nPDF constraints from virtual photons}\label{sec:npdfs}

Closely related to prompt photons is the Drell-Yan (DY) process  
whereby the invariant mass ($M$) of the photon is non-zero. 
At leading order, $q \bar{q} \to \gamma^\star$, DY is insensitive to FCEL 
due to the absence of a recoiling parton~\cite{Arleo:2015qiv}. 
Real emissions start at NLO, 
leading to 
$\Delta E_\textnormal{FCEL} \propto Q_s / \sqrt{M^2 + \pt^2} \ll 1\,$, 
and consequently the role of nPDFs prevails.  
Low-mass DY in pA collisions at the LHC 
would thus be an ideal probe of nuclear shadowing at small-$x\,$, 
being largely uncontaminated by energy loss effects like FCEL. 
We demonstrate this explicitly by generating pseudo-data 
for $\RpPb(y)$
at the collisional energy of 
$\sqrt{s_\textnormal{NN}} = 8.8$~TeV, 
and apply reweighting techniques to existing 
nPDF sets. 
(A similar demonstration was made for 
future EIC measurements in Ref.~\cite{Armesto:2023hnw}.) 

First let us describe our recipe for producing the pseudo-data, 
using a setup in accordance with Runs 3 and 4 at 
LHCb~\cite{LHCb-CONF-2018-005}. 
Here we focus on the range
$5 < M < 9$~GeV, 
and $p_{_\perp}$-integrated result 
(to avoid complications such as the Cronin effect). 
To simplify notation, we refer to the 
cross-section as rapidity differential, i.e. 
$\sigma_i \equiv {\rm d}\sigma_i/{\rm d}y$ where
$i$ denotes the nPDF member set and $j$ will be reserved 
for the relevant rapidity bin $y_j$, 
$\sigma_i(j) = {\rm d}\sigma_i(y_j)/{\rm d}y\,$.
We consider the following rapidity bins,\footnote{%
  Strictly speaking, the cross-section for bin $y_j$ 
  ($j=1,...,8$) is given 
  by $\frac1{\Delta y} \int_{y_\textnormal{min}}^{y_\textnormal{max}} 
  {\rm d}y \frac{{\rm d}\sigma(y)}{{\rm d}y}\,$. 
}
\bea
  \text{backward:} & & 
  [-5, -4.5] 
  \; , \ 
  [-4.5, -4] 
  \; , \ 
  [-4, -3.5] 
  \; , \ 
  [-3.5, -3]  
  \; ;
  \nonumber
  \\[0mm]
  \text{forward:} & & 
  [+2, +2.5] 
  \; , \ 
  [+2.5, +3] 
  \; , \ 
  [+3, +3.5] 
  \; , \ 
  [+3.5, +4]  
  \; .
\eea
A perturbative prediction for 
$\sigma_i^\textnormal{pp}(j)$ and 
$\sigma_i^\textnormal{pPb}(j)$ 
can be obtained with the \texttt{DYTurbo} code~\cite{Camarda:2019zyx}~(adapted
for asymmetric collisions). 
We do so at NLO, with 
factorization and renormalization scales $\mu^2 = {\cal O}(M^2)$, 
and focus on the sets 
provided\footnote{%
  Specifically, we use those sets which do not use LHCb $D$-meson 
  data and having $N_\textnormal{rep}=250$ replicas. 
} by nNNPDF3.0~\cite{AbdulKhalek:2022fyi}. 
We denote the corresponding $\RpA$ 
from \eq{eq:RpA} by $R(j)$, 
for which the statistical error $\delta R$ can be estimated 
from 
\bea
  \frac{\delta R(j)}{R(j)}
  & = &
  \sqrt{
    \left[
    \frac{1}{N_{i=0}^\textnormal{pPb}(j)}
    +
    \frac{1}{N_{i=0}^\textnormal{pp}(j)}
  \right]
  \left[
    1 + \frac{1}{(S/B)_{j, \text{eff}}} 
    \right]
  }  
  \; , \,
  \la{uncertainty}
\eea
where we use the signal-to-background 
ratio 
${(S/B)_{j, \text{eff}}} \simeq 1/30$, 
as compatible with 
Ref.~\cite{LHCb-CONF-2018-005}. 
From the cross-sections
(output by \texttt{DYTurbo}), one may then estimate 
\bea
  N_i^\textnormal{pp} (j) 
  \; = \; 
  {\cal A} \epsilon
  \times 
  {\cal L}_\textnormal{pp}
  \times
  \sigma_i^\textnormal{pp} (j) 
  \; ;
  & &
  N_i^\textnormal{pPb} (j) 
  \; = \;
  {\cal A} \epsilon
  \times 
  {\cal L}_\textnormal{pPb}
  \times
  \sigma_i^\textnormal{pPb} (j) 
  \; , 
  \la{NpA} 
\eea
where acceptance times efficiency 
${\cal A} \epsilon \approx 0.9$
and foreseen luminosities of 
${\cal L}_\textnormal{pp} = 104~\text{pb}^{-1}$ 
and 
${\cal L}_\textnormal{pPb} \approx 250~\text{nb}^{-1}$. 
With these values, we obtain 
a relative uncertainty of $\sim 5$\%
in \eq{uncertainty}. 
Finally, we sampled data from a normal 
distribution (${\cal N}$) 
centered on $\hat R$ and with statistical variance $\delta R^2$\,:
\be
  R_\textnormal{data}(j) 
  \ \sim \ 
  {\cal N}\big(
  \hat R(j) , 
  \delta R^2(j)
  \big) \, ,
\ee
where the mean value is given by 
the central nPDF set 
$\hat R(j) \equiv R_{\pA}^\textnormal{{nNNPDF}}(j)\big|_{i=0}\,$.

Next, we apply the reweighting method  
using the Bayesian approach 
(as applicable to NNPDF sets) 
which is briefly summarised here. 
(A more thorough discussion of both Bayesian and Hessian methods 
can be found, e.g. in Ref.~\cite{Paukkunen:2014zia}.) 
Given a large ensemble of of (n)PDFs $f_i\,$, $i=1 ... N_\textnormal{rep}\,$, 
sampled with equal weights
from the underlying functional probability distribution, 
the expectation value and variance of an observable may be computed from 
\be
  \textstyle
  \langle {\mathbb O} \rangle_\textnormal{old}
  \; =\;
  \frac{1}{N_\textnormal{rep}} \sum_{i=0}^{N_\textnormal{rep}} {\mathbb O}[f_i]
  \; , \qquad
  \delta \langle {\mathbb O} \rangle_\textnormal{old}
  \; =\;
  \sqrt{
  \frac{1}{N_\textnormal{rep}} \sum_{i=0}^{N_\textnormal{rep}} 
  \big( {\mathbb O}[f_i] - \langle {\mathbb O} \rangle_\textnormal{old} \big)^2
  }
  \; .
  \la{old}
\ee
Given new measurements, the $\chi^2_i$ for each replica can be calculated, 
which gives a notion of ``how far'' that replica is from reproducing the data.  
If the data are uncorrelated, 
\be
  \textstyle
  \chi^2_i
  \; = \;
  \sum_{j=1}^{N_\textnormal{data}} 
  \Big[ 
  \frac{ R_\textnormal{data}(j) - R_\textnormal{th}^i(j) }{\delta R_\textnormal{data} (j)} 
  \Big]^2 \, ,
\ee
for which each $i$ is assigned the weight
\be
  \textstyle
  w_i 
  \; = \; 
  \frac1{\cal D} \,
  \big( \chi^2_i \big)^{(N_\textnormal{data}-1)/2} 
  e^{ - \chi^2_i/2 }
  \, ;
  \qquad
  {\cal D} \; \equiv \;
  \sum_{i=1}^{N_\textnormal{data}}\big( \chi^2_i \big)^{(N_\textnormal{data}-1)/2} 
  e^{ - \chi^2_i/2 }
  \, ,
\ee
where the normalisation constant ${\cal D}$ ensures that $\sum_i w_i = 1\,$. 
This gives us ``reweighted'' 
predictions for a given quantity, namely
\be
  \textstyle
  \langle {\mathbb O} \rangle_\textnormal{new}
  \; =\;
  \sum_{i=0}^{N_\textnormal{rep}} w_i \, {\mathbb O}[f_i]
  \; , \qquad
  \delta \langle {\mathbb O} \rangle_\textnormal{new}
  \; =\;
  \sqrt{
  \sum_{i=0}^{N_\textnormal{rep}} 
  w_i \, \big( {\mathbb O}[f_i] - \langle {\mathbb O} \rangle_\textnormal{new} \big)^2
  }
  \; .
  \la{new}
\ee

\begin{figure}[t]
\centering{
\hspace{-9mm}
\includegraphics[scale=0.65]{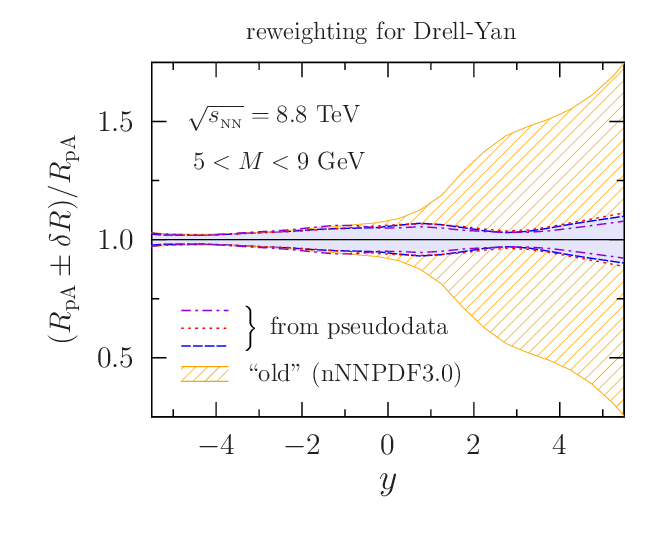}
\hspace{0mm}
\includegraphics[scale=0.65]{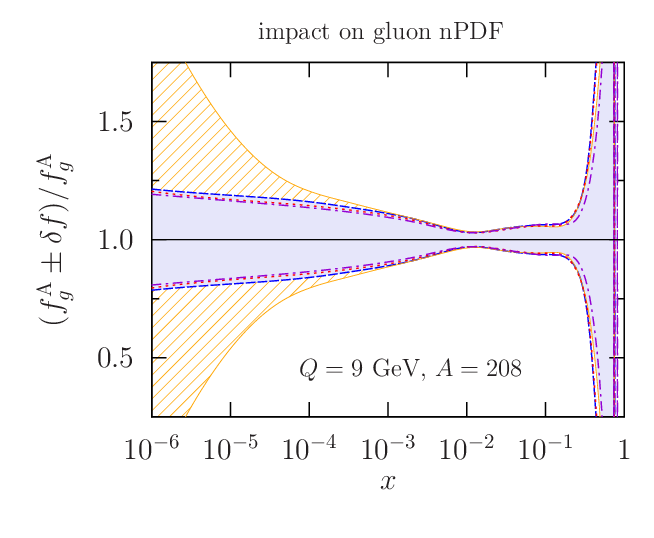}}
\vspace{-5mm}
\caption{\la{fig:reweighted}
  Left: relative error 
  in the predicted $R^\textnormal{DY}_{\pA}$ 
  as a function of rapidity, before and after reweighting. 
  Right: the same, but for the 
  gluon nPDF $f^\textnormal{A}_g(x,Q=9~{\rm GeV})$ 
  as a function of $x\,$. 
    Note that in both panels, only the uncertainty in the nPDFs 
  is shown and calculated according to \eq{old} and \eq{new}. 
}
\end{figure}

The observable corresponding to the pseudo-data, 
${\mathbb O} = R^\textnormal{DY}_\textnormal{pPb}$, is 
displayed in Figure~\ref{fig:reweighted}. 
We generate three independent sets of pseudo-data, 
which each lead to slightly varying predictions for 
the modification factor. 
However, the relative improvement in the error\footnote{%
  We recognize that Ref.~\cite{AbdulKhalek:2022fyi} 
  advocates computing the error with confidence level intervals 
  rather than the variance  (see Sec.~7.2 of that work). 
  However, we use \eq{new} for simplicity and 
  because our purpose here is not to make actual 
  predictions but to motivate future measurements. 
}
is similar in all cases with and especially significant at 
forward rapidity (by construction). 
The consequence of these reweightings for the gluon nPDF 
is also shown, demonstrating the constraining 
power at small $x$~(and similarly for the quarks, not shown). 
Although slightly less stringent than results obtained with $D$-mesons, 
it may be more reliable as DY is comparatively unaffected by FCEL.

\section{Conclusion}

In summary, 
we have estimated influence of fully coherent energy loss and energy gain on prompt photon 
production in pA collisions. 
Although these effects are generally small, they 
are not negligible 
(especially at low $\pt$ and backward rapidity $y<0$), 
and arise from a subtle interplay between 
energy loss and energy gain mechanisms. 
On the other hand, the DY process 
remains largely impervious to medium induced coherent gluon radiation, 
making it one of the most promising probes of nuclear PDFs. 
Using realistic run 3 pseudo-data, we performed an NLO reweighting 
study and found that DY measurements 
can provide significant constraints on both quark and gluon 
densities at small-$x$.

\acknowledgments
{\small
We thank Jean-Philippe Guillet for discussions and Michael Winn for discussions and for providing us with the acceptance/efficiency and luminosity expected with LHCb at Run~3. G.J. is supported by the Agence Nationale de la Recherche (ANR) under grant ANR-22-CE31-0018.
}

\providecommand{\href}[2]{#2}\begingroup\raggedright\endgroup

\begin{thebibliography}{10}

\bibitem{Arleo:2010rb}
F.~Arleo, S.~Peign{\'e} and T.~Sami, \emph{{Revisiting scaling properties of
  medium-induced gluon radiation}},
  \href{https://doi.org/10.1103/PhysRevD.83.114036}{\emph{Phys. Rev.}
  {\bfseries D83} (2011) 114036}
  [\href{https://arxiv.org/abs/1006.0818}{{\ttfamily 1006.0818}}].

\bibitem{Arleo:2012hn}
F.~Arleo and S.~Peign{\'e}, \emph{{J/$\psi$ suppression in pA collisions from
  parton energy loss in cold QCD matter}},
  \href{https://doi.org/10.1103/PhysRevLett.109.122301}{\emph{Phys. Rev. Lett.}
  {\bfseries 109} (2012) 122301}
  [\href{https://arxiv.org/abs/1204.4609}{{\ttfamily 1204.4609}}].

\bibitem{Arleo:2020eia}
F.~Arleo and S.~Peign{\'e}, \emph{{Quenching of light hadron spectra in pA
  collisions from fully coherent energy loss}},
  \href{https://doi.org/10.1103/PhysRevLett.125.032301}{\emph{Phys. Rev. Lett.}
  {\bfseries 125} (2020) 032301}
  [\href{https://arxiv.org/abs/2003.01987}{{\ttfamily 2003.01987}}].

\bibitem{Arleo:2020hat}
F.~Arleo, F.~Cougoulic and S.~Peign\'e, \emph{{Fully coherent energy loss
  effects on light hadron production in pA collisions}},
  \href{https://doi.org/10.1007/JHEP09(2020)190}{\emph{JHEP} {\bfseries 09}
  (2020) 190} [\href{https://arxiv.org/abs/2003.06337}{{\ttfamily
  2003.06337}}].

\bibitem{Arleo:2021bpv}
F.~Arleo, G.~Jackson and S.~Peign{\'e}, \emph{{Impact of fully coherent energy
  loss on heavy meson production in pA collisions}},
  \href{https://doi.org/10.1007/JHEP01(2022)164}{\emph{JHEP} {\bfseries 01}
  (2022) 164} [\href{https://arxiv.org/abs/2107.05871}{{\ttfamily
  2107.05871}}].

\bibitem{Arleo:2021krm}
F.~Arleo, G.~Jackson and S.~Peign{\'e}, \emph{{Depletion of atmospheric
  neutrino fluxes from parton energy loss}},
  \href{https://doi.org/10.1016/j.physletb.2022.137541}{\emph{Phys. Lett. B}
  {\bfseries 835} (2022) 137541}
  [\href{https://arxiv.org/abs/2112.10791}{{\ttfamily 2112.10791}}].

\bibitem{Eskola:2021nhw}
K.J.~Eskola, P.~Paakkinen, H.~Paukkunen and C.A.~Salgado, \emph{{{EPPS21}}:
  {{A}} global {{QCD}} analysis of nuclear {{PDFs}}},
  \href{https://doi.org/10.1140/epjc/s10052-022-10359-0}{\emph{The European
  Physical Journal C} {\bfseries 82} (2022) 413}
  [\href{https://arxiv.org/abs/2112.12462}{{\ttfamily 2112.12462}}].

\bibitem{AbdulKhalek:2022fyi}
R.~Abdul~Khalek, R.~Gauld, T.~Giani, E.R.~Nocera, T.R.~Rabemananjara and
  J.~Rojo, \emph{{nNNPDF3.0: evidence for a modified partonic structure in
  heavy nuclei}},
  \href{https://doi.org/10.1140/epjc/s10052-022-10417-7}{\emph{Eur. Phys. J. C}
  {\bfseries 82} (2022) 507}
  [\href{https://arxiv.org/abs/2201.12363}{{\ttfamily 2201.12363}}].

\bibitem{Duwentaster:2022kpv}
P.~Duwent{\"a}ster, T.~Je{\v z}o, M.~Klasen, K.~Kova{\v r}{\'i}k, A.~Kusina,
  K.F.~Muzakka et~al., \emph{Impact of heavy quark and quarkonium data on
  nuclear gluon {{PDFs}}},
  \href{https://doi.org/10.1103/PhysRevD.105.114043}{\emph{Phys. Rev. D}
  {\bfseries 105} (2022) 114043}.

\bibitem{Arleo:2025oos}
F.~Arleo et~al., \emph{{Nuclear Cold QCD: Review and Future Strategy}},
  \href{https://arxiv.org/abs/2506.17454}{{\ttfamily 2506.17454}}.

\bibitem{Arleo:2007js}
F.~Arleo and T.~Gousset, \emph{{Measuring gluon shadowing with prompt photons
  at RHIC and LHC}},
  \href{https://doi.org/10.1016/j.physletb.2007.12.025}{\emph{Phys. Lett.}
  {\bfseries B660} (2008) 181}
  [\href{https://arxiv.org/abs/0707.2944}{{\ttfamily 0707.2944}}].

\bibitem{Arleo:2011gc}
F.~Arleo, K.J.~Eskola, H.~Paukkunen and C.A.~Salgado, \emph{{Inclusive prompt
  photon production in nuclear collisions at RHIC and LHC}},
  \href{https://doi.org/10.1007/JHEP04(2011)055}{\emph{JHEP} {\bfseries 04}
  (2011) 055} [\href{https://arxiv.org/abs/1103.1471}{{\ttfamily 1103.1471}}].

\bibitem{Helenius:2014qla}
I.~Helenius, K.J.~Eskola and H.~Paukkunen, \emph{{Probing the small-$x$ nuclear
  gluon distributions with isolated photons at forward rapidities in pPb
  collisions at the LHC}},  \href{https://arxiv.org/abs/1406.1689}{{\ttfamily
  1406.1689}}.

\bibitem{Arleo:2015qiv}
F.~Arleo and S.~Peign{\'e}, \emph{{Disentangling Shadowing from Coherent Energy
  Loss using the Drell-Yan Process}},
  \href{https://doi.org/10.1103/PhysRevD.95.011502}{\emph{Phys. Rev.}
  {\bfseries D95} (2017) 011502}
  [\href{https://arxiv.org/abs/1512.01794}{{\ttfamily 1512.01794}}].

\bibitem{Aurenche:2006vj}
P.~Aurenche, M.~Fontannaz, J.-P.~Guillet, {\'E}.~Pilon and M.~Werlen, \emph{A
  new critical study of photon production in hadronic collisions}, {\emph{Phys.
  Rev.} {\bfseries D73} (2006) 094007}
  [\href{https://arxiv.org/abs/hep-ph/0602133}{{\ttfamily hep-ph/0602133}}].

\bibitem{Arleo:2012rs}
F.~Arleo and S.~Peign{\'e}, \emph{{Heavy-quarkonium suppression in pA
  collisions from parton energy loss in cold QCD matter}},
  \href{https://doi.org/10.1007/JHEP03(2013)122}{\emph{JHEP} {\bfseries 03}
  (2013) 122} [\href{https://arxiv.org/abs/1212.0434}{{\ttfamily 1212.0434}}].

\bibitem{Jackson:2023adv}
G.~Jackson, S.~Peign{\'e} and K.~Watanabe, \emph{{Coherent gluon radiation:
  beyond leading-log accuracy}},
  \href{https://doi.org/10.1007/JHEP05(2024)207}{\emph{JHEP} {\bfseries 05}
  (2024) 207} [\href{https://arxiv.org/abs/2312.11650}{{\ttfamily
  2312.11650}}].

\bibitem{in-prep}
F.~Arleo, D.~Bourgeais and G.~Jackson, in preparation.

\bibitem{Armesto:2023hnw}
N.~Armesto, T.~Cridge, F.~Giuli, L.~Harland-Lang, P.~Newman, B.~Schmookler
  et~al., \emph{{Impact of inclusive electron ion collider data on collinear
  parton distributions}},
  \href{https://doi.org/10.1103/PhysRevD.109.054019}{\emph{Phys. Rev. D}
  {\bfseries 109} (2024) 054019}
  [\href{https://arxiv.org/abs/2309.11269}{{\ttfamily 2309.11269}}].

\bibitem{LHCb-CONF-2018-005}
M.~Winn, ``{LHCb projections for proton-lead collisions during LHC Runs 3 and
  4}.'' \url{http://cds.cern.ch/record/2648625}.

\bibitem{Camarda:2019zyx}
S.~Camarda et~al., \emph{{DYTurbo: Fast predictions for Drell-Yan processes}},
  \href{https://doi.org/10.1140/epjc/s10052-020-7757-5}{\emph{Eur. Phys. J. C}
  {\bfseries 80} (2020) 251}
  [\href{https://arxiv.org/abs/1910.07049}{{\ttfamily 1910.07049}}].

\bibitem{Paukkunen:2014zia}
H.~Paukkunen and P.~Zurita, \emph{{PDF reweighting in the Hessian matrix
  approach}}, \href{https://doi.org/10.1007/JHEP12(2014)100}{\emph{JHEP}
  {\bfseries 12} (2014) 100} [\href{https://arxiv.org/abs/1402.6623}{{\ttfamily
  1402.6623}}].

\end{thebibliography}

\end{document}